\documentstyle[eqsecnum,aps,epsf]{revtex}
\draft
\def\FIG#1{\epsfxsize=\columnwidth\epsfbox{#1}}
\begin{document}
\title{Black holes and gravitational waves:
simultaneous discovery by
initial laser interferometers}

\author{V.~M.~Lipunov, K.~A.~Postnov, and M.~E.~Prokhorov}
\address{Moscow State University, Physical Department,
Sternberg Astronomical Institute, 119899 Moscow, Russia
e-mail: lipunov@sai.msu.su; pk@sai.msu.su; mike@sai.msu.su\\
fax: +7 (095) 932 88 41}
\date{\today}
\maketitle
\begin{abstract}

Study of gravitational-radiation induced merging rates of relativistic
binary stars (double neutron stars; neutron star + black hole; double
black holes) shows that the first-generation gravitational wave
interferometers with an rms-sensitivity of $10^{-21}$ at frequency
$100$ Hz can detect 10-700 black hole and only $\sim 1$ neutron star
coalescences in a 1-year integration time in a wide range of stellar
evolution parameters.  It is notable that modern concepts of stellar
evolution predict that the first detection of gravitational wave will
independently discover black holes.

\end{abstract}


\narrowtext

The final merging stage of binary relativistic star evolution
containing two compact stars (neutron stars (NS) or black holes (BH))
that merge due to gravitational-radiation induced orbit decay on a
time-scale shorter than the Hubble time ($1.5\times 10^{10}$ years) are
among the primary targets for gravitational wave (GW) interferometers
currently under construction (LIGO, VIRGO, GEO-600) \cite{GW-int}.
Reality of such events in the Universe is confirmed by binary pulsar
observations \cite{Hulse}.  To date, 5 binary NS are known in our
Galaxy; 3 of them should coalesce within the Hubble time \cite{5PSR}.
Based on the properties of these binary pulsars, one may estimate the
total number of such systems and hence the merging rate of binary NS in
the Galaxy, and then extrapolate this rate to the volume of the
Universe where GW-detectors will be able to detect GW-signal at a given
signal-to-noise level.  Such estimations made in the last five years
\cite{Phinney} yield a ``realistic'' galactic rate of double NS mergings
of $10^{-5}-10^{-6}$ per year.

Theoretical estimate of double NS merging rate may be deduced from
stellar evolution \cite{Th_rate} and give $\approx 10^{-4}$ NS-NS
coalescences per year per Galaxy. It can be shown that these two
estimates do not contradict to each other if one takes into account
NS+NS binaries which contain no pulsar (ejecting NS) and thus
are unobservable by traditional radioastronomical means
\cite{PSP_deficit}.

The stellar evolution theory also predicts that neutron star -- black
hole (NS+BH) and black hole -- black hole (BH+BH) binaries should be
present in the Galaxy. So far the indirect existence of BH in stellar
binary systems has been argumented by a high mass ($>3 M_\odot$) of the
unseen companion in 11 close X-ray binary systems \cite{BH_cand}.  No
PSR+BH binary system has yet been discovered in spite of optimistic
theoretical expectations \cite{PSR+BH} of $\sim$ 1 per 1000 single
pulsars. Theoretical estimates of NS+BH/BH+BH merging rate in the
Galaxy are 1-2 orders of magnitude lower than NS+NS merging rate. The
difficulty in doing such estimates is that additional parameters of BH
formation appears: the critical mass $M_{cr}$ beyond which the star
collapses into BH and the mass of BH formed.

From the point of view of GW detection, NS+BH and BH+BH binary systems
are advantageous in having higher "chirp mass" (${\cal
M}=M(\mu/M)^{3/5}$, $M=M_1+M_2$, $\mu=M_1M_2/M$) that determines the
amplitude of the dimensionless metric strain produced by a binary
system:  $h_{amp}\sim {\cal M}^{5/3} f^{2/3}/r$, where $f$ is the GW
frequency and $r$ is the distance to the binary.  In fact, matched
filtering technique of data analysis \cite{300yr} allows enhancement of
the signal-to-noise ratio from coalescing binaries $S/N\propto
\sqrt{n}$ where $n=f^2/\dot f$ is the number of cycles of the signal
passing through the detector frequency band $\Delta f\sim f$, so one
usually considers the characteristic strain amplitude
$h_c=h_{amp}\sqrt{n} \sim {\cal M}^{5/6} f^{-1/6}/r$.  This means that
at a given $S/N$ the detector will be sensitive to a distance $r\propto
{\cal M}^{5/6}$. For example, if we consider a NS+NS binary consisting
of two identical NS $M_1=M_2=1.4 M_\odot$ (${\cal M}_{ns}\approx 1.2
M_\odot$) and a BH+BH binary consisting of two identical BH $M_1=M_2=10
M_\odot$ (${\cal M}_{bh}\approx 8.7 M_\odot$), the limiting distance
$r_{bh}\simeq 5.2 r_{ns}$, or the volume available to search for such
BH+BH systems is 140 times as large as the volume in which NS+NS
binaries can be detected. Assuming homogeneous matter distribution, the
detection rate in a 1-year integration time with a given signal-to-noise
ratio is
\[
N_{bh}/N_{ns}=(R_{bh}/R_{ns})\times({\cal M}_{bh}/{\cal
M}_{ns})^{15/6}
\]
where $R_{bh}$ and $R_{ns}$ are galactic merging rates for BH and NS
binaries, respectively.  Clearly, to answer the question what type of
binary systems will be more numerous during 1-year operation time of a
GW-interferometer one should study in detail different types of binary
mergings in the Galaxy.

In this paper we present calculations of NS+NS, NS+BH and BH+BH
galactic merging rates in a wide range of main parameters of stellar
evolution. We address the question  how many binary mergings can one
expect to detect with the rms-sensitivity $h_{rms}\simeq 10^{-21}$ at
$f_c=100$ Hz corresponding to $S/N=1$ for GEO-600 and $S/N=3$ for
LIGO/VIRGO-I \cite{GW-int}.  We find that the prospect to discover
BH-events (binary BH merging or BH+NS merging) is at least 10 times better
than for NS+NS mergings. Moreover, for currently
popular high recoil velocities of young pulsars ($\sim 400 $ km/s;
\cite{L&L}) only BH+BH binaries may be detected by the first-order
laser interferometers.

Qualitatively, a crude estimate of the ratio of BH-containing binary
merging rate to NS binary merging rate may be done as follows.  BH is
thought to result from the core collapse of massive stars. Single
massive stars lose roughly half their mass through intensive stellar
wind; when in a close binary system, the mass may be transferred onto
the secondary companion. An approximate relation between the initial
mass of the star and its core is $M_{core}\simeq 0.1 M_{ms}^{1.4}$
\cite{evol}.  Let us assume the mass of BH progenitor just before the
collapse be $35 M_\odot$, which would correspond roughly to
$M_{ms}\sim 60 M_\odot$ on the initial main sequence.  On the other
hand, any star with $M_{ms}\ge 10 M_\odot$ evolves to form a NS.
Using the Salpeter mass function for star formation ($dN/dM \approx
1\,(M/M_\odot)^{-2.35}$ star per year), we obtain that BH formation
rate relates to NS formation rate as $(60/10)^{-1.35}\approx 0.09$.
Extrapolating this logic to binary BH/NS systems, we might expect
$R_{bh}/R_{ns}\sim 1/10$, to a half-order accuracy. Actually, the
situation is complicated by several factors: an asymmetry of the
supernova explosion which may act more efficiently in the case of NS
formation; mass exchange between the components; distribution by mass
ratio, etc. All these factors will be accounted for in our
calculations.

To perform evolutionary calculations, we employ Monte-Carlo method for
binary stellar evolution studies developed by us over last ten years
(the Scenario Machine code); we refer to \cite{SM} for a detailed
description of the method and evolutionary scenarios used.  The basic
idea is to calculate the evolution of a representative number of binary
stars (typically $10^6$) whose orbital and physical parameters are
distributed corresponding to the observational data when available
(e.g. initial masses, semimajor axes, recoil velocities of newborn
neutron stars), or are chosen according to some model laws (e.g.
initial rotation periods and magnetic fields of neutron stars).  The
calculated number of mergers is then scaled to the total stellar mass
of the Galaxy (taken $10^{11} M_\odot$) and after dividing by the age
of the Galaxy ($1.5\times 10^{10}$ years) yields the galactic merging
rate $R$. Note (see below) that when reducing to the total mass
of the Galaxy we must take into account the fraction of stars entering
binary systems. This fraction is at least $50\%$

A BH is known to be fully described by three parameters: its mass
$M_{bh}$, angular momentum, and electric charge. For our purposes,
however, only mass is important as it determines the orbital evolution
when the BH resides in a close binary system.  At present, stellar-mass
BH are thought to result from the core collapse of high-mass stars with
$M>M_{*}$ where $M_*$ is the pre-supernova mass (see \cite{snowmass}
for recent discussion).  It seems reasonable to assume that the mass of
BH is proportional to $M_*$, i.e.  $M_{bh}=k_{bh}M_*$, $0<k_{bh}\le 1$.
This mass in turn is calculated from modern stellar evolution theory.
Analysis of observational data on BH-candidates in binary systems
\cite{PSR+BH} shows that the most plausible BH-formation parameters are
$k_{bh}=0.25-0.5$, $M_{*}=25-50$. Some restriction to these parameters
are discussed below.

When in binaries, another important parameter appears:  the
additional velocity $w_{bh}$ imparted to BH during the anisotropic
collapse.  Here several different possibilities are feasible: (1)
$w_{bh}$ is proportional to the mass ejected during the collapse
$M_{ej}= M_*(1-k_{bh})$; (2) the momentum $M_{bh}w_{bh}$ is
proportional to the ejected mass $M_{ej}$; (3) $M_{bh}w_{bh}$ is
proportional to the ejected momentum.

We shall assume a universal mechanism giving anisotropic velocity for
both NS and BH. In the case (1) we have:
\begin{equation}
w_{bh}=w_{ns}(1-k_{bh})(1-M_{OV}/M_*)^{-1}
\label{wbh1}
\end{equation}
where $M_{OV}=2.5$ M$_\odot$ is the Oppenheimer-Volkoff limit for NS
mass. This law is chosen  assuming boundary conditions $w_{bh}=0$ at
$k_{bh}=1$ (i.e. when the total mass of the collapsing star goes into a
BH) and $w_{bh}=w_{ns}$ once $M_{bh}=M_{OV}$.  Clearly, the case (2) is
similar to the case (1) but produces lower BH-velocities while the case
(3), instead, assumes $w_{bh}=w_{ns}$ irrespective of BH mass and thus
produces higher BH velocities.  We found, however, that all three cases
yield BH+BH/BH+NS merging rates practically coinciding with each other
at high kick velocities, so here we represent the results for the case
(1) only.

The three-dimensional NS kick velocity is assumed to be arbitrarily
directed in space and to be distributed so as to reproduce the observed
pulsars' transverse (two-dimensional) velocities \cite{L&L} (see
\cite{SM} for more detail):
\begin{equation}
f_{LL}(x)dx\propto x^{0.19}(1+x^{6.72})^{-1/2}dx
\label{LLkick}
\end{equation}
where $x=w/w_0$, $w_0$ is the characteristic velocity; the observed
Lyne-Lorimer 2D-distribution is obtained at $w_0=400$~ km/s. The mean
3-D kick velocity differ from $w_0$ by $\approx 10\%$.  Our analysis of
modern evolutionary scenarios \cite{SM} shows that the most probable
kick velocities lie in the range 200-400 km/s.

\begin{figure}
\FIG{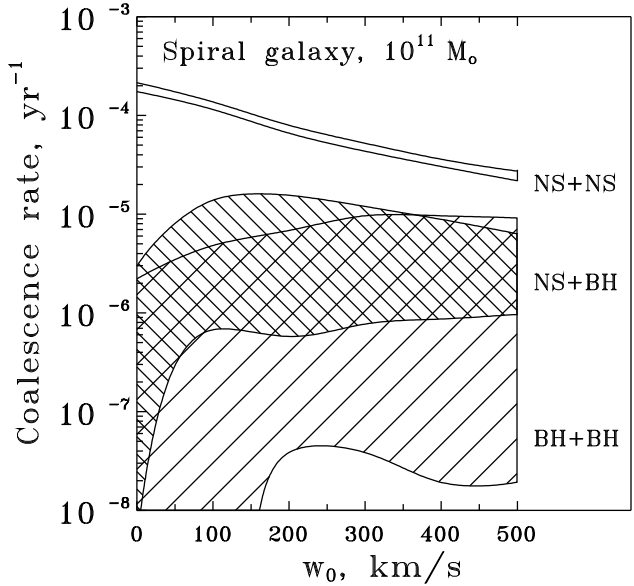}
\caption{The merging rate of NS+NS, NS+BH, and BH+BH binaries for
Lyne-Lorimer kick velocity distribution as a function of $w_0$ assuming
BH formation parameters $M_*=15-50$ M$_\odot$, $k_{bh}=0.25$, for
different scenarios of binary star evolution in a model spiral galaxy
with the total stellar mass $10^{11}$ M$_\odot$ and $50\%$ baryon
fraction in binary stars.}
\label{rate}
\end{figure}

In Fig. \ref{rate} we plot the relativistic compact binaries' merging
rates as a function of the mean kick velocity assuming Lyne-Lorimer
distribution (\ref{LLkick}). BH-formation parameters were taken within
the limits $M_*=15-50 M_\odot$, $k_{bh}=0.25$.  Different scenarios for
stellar evolution were considered (see \ref{PSR_deficit} for more
detail about evolutionary scenarios).  From Fig.  \ref{rate} we see
that the theoretical expectation for the NS+NS merging rate in a model
spiral galaxy with typical mass of $0.5\times 10^{11}$ M$_\odot$ in binary
stars lie within the range from $\sim 3\times 10^{-4}$ yr$^{-1}$ to
$\sim 3\times 10^{-5}$ yr$^{-1}$, depending on the assumed mean kick
velocity and the shape of its distribution. Note that 
for spherically symmetric collapse the calculated rates 
coincide well with those found by Tutukov \&
Yungelson \cite{Th_rate} who used another method
of calculations. For Lyne-Lorimer law with
the mean value of 400 km/s, we obtain $R_{NS+NS}\approx 5\times
10^{-5}$ yr$^{-1}$. BH+BH and BH+NS merging rates are typically 1-2
orders of magnitude smaller.

Two details in Fig. \ref{rate} are worth noting: 1) the binding effect
at small kick velocities and 2) the smaller effect of high kicks on the
BH+BH/BH+NS rate.  The first fact is qualitatively clear: a high kick
leads to the system disruption; however, if the system has survived the
explosion, its orbit would have a periastron always smaller than in the
case without kick. During the subsequent tidal circularization a closer
binary system will form which will spend less time prior to the merging.
The binding effect at small recoil velocities is more pronounced in the
case of binary BH. At higher kicks their merging rate decreases slower
than that of binary NS due to higher masses of the components.

Having found galactic merging rates $R$, we can calculate them within a
volume accessible for GW-observations of each type of binaries (which
from the point of view of GW-signal differ, to the first approximation,
in having different chirp masses). Here we may either use normalization
to the IR-luminosity per cubic megaparsec (as in Phinney's paper
\cite{Phinney}), or scale the galactic rates per baryonic density in
stars and then extrapolate it to the volume desired. In the latter case
we obtain the formula
\[
{\cal R} = 0.0063\,R\,(\epsilon_d/0.5)
\left(\Omega_b/0.0046\right)h_{75}^2\hbox{Mpc}^{-3}
\]
where $\Omega_b$ is the baryon density (in units of critical density to
close the Universe), $h_{75}=H_0/75$ km/s/Mpc is the present value of
the Hubble constant, $\epsilon_d$ is the fraction of baryons contained
in binary stars. Typically $0.25<\epsilon_d<0.75$.  This normalization
coincides with Phinney's one assuming $\epsilon_d\Omega_b=0.0046$.
According to \cite{Peebles}, $\Omega_b=0.0015$ in stars within galactic
disks and $\Omega_b=0.003$ in spherical bulges of spiral galaxies and
in elliptical galaxies.  Calculations of NS+NS merging rate in
elliptical galaxies show \cite{Green_f} that although a strong decrease
in binary merging rate occurs, it tends to a nearly constant value
several times lower than the binary merging rate in a spiral galaxy
with constant star formation. So taking $\epsilon_d\Omega_b=0.5\times
0.0046$ would not be too far off and we use it to evaluate binary
merging rates in a given volume of the Universe.  Note, however, that
in view of unknown exact value of $\epsilon_d$ all absolute rates we
obtain may be uncertain to within a factor of 2.

\begin{figure} 
\FIG{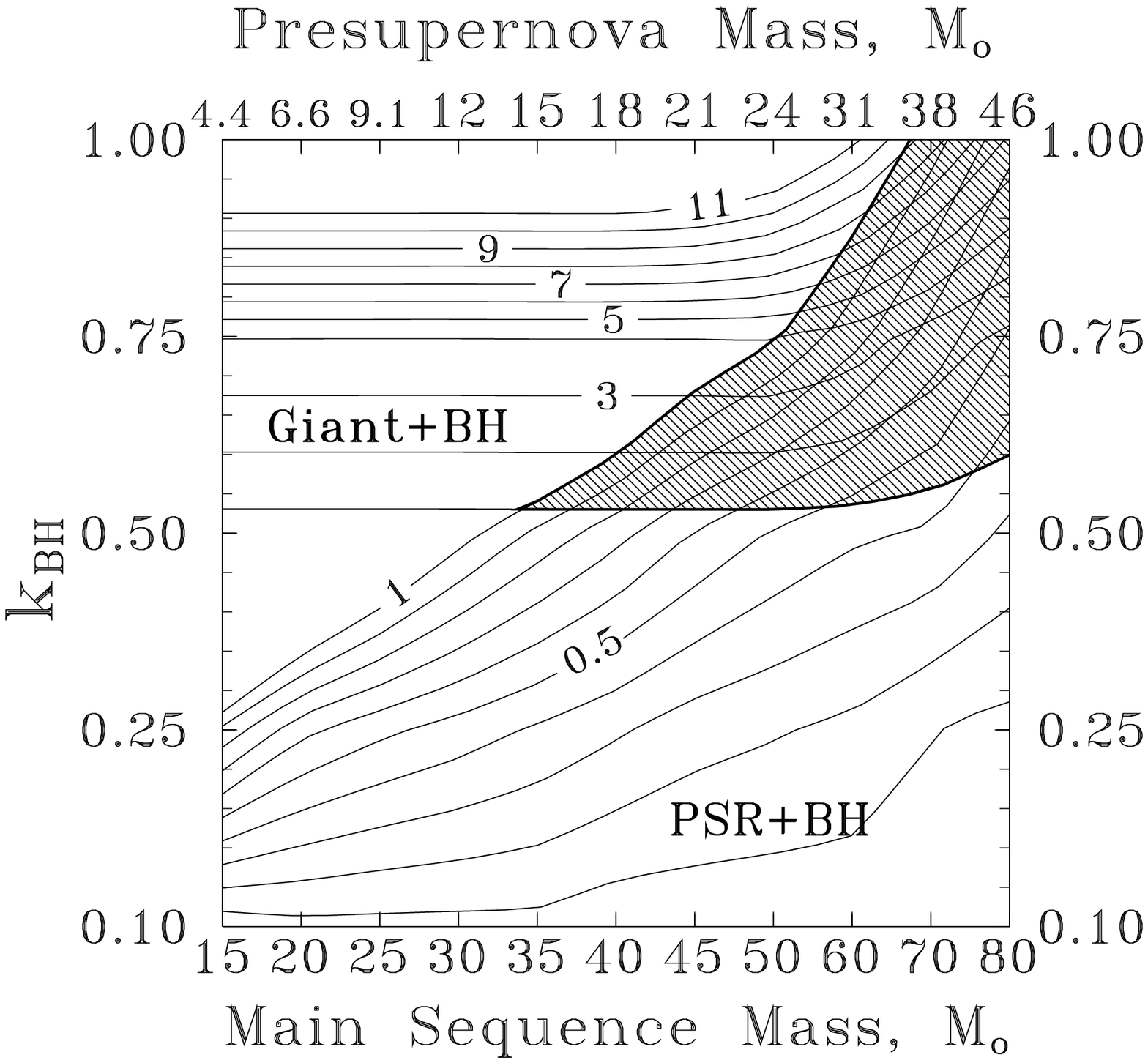} 
\caption{The galactic number of Cyg
X-1-like systems (upper curves) and the fraction of binary PSR with BH
among the total number of single PSR in the Galaxy (lower curves) as a
function of BH formation parameters. The upper horizontal scale is in
presupernova masses $M_*$, the lower scale in the critical main
sequence mass $M_{cr}$ (both in solar masses). The fraction of
presupernova mass to collapse into BH $k_{bh}$ is along the vertical
scale. The shaded region shows the most plasuible BH formation
parameters.}
\label{cyg_psr}
\end{figure}

Consider now whether is it possible to restrict 
BH formation parameters from the existing astronomical 
observations. From known 11 BH candidates in binary systems
the evolutionary status of Cyg X-1 is mostly well understood
(a massive X-ray binary containing a 10 $M_\odot$ BH and 
a 30 $M_\odot$ OB-supergiant). The evolution of 6 X-ray Novae -- 
BH candidates is much more controversial. Only one CYg X-1-like source
is observed in the Galaxy, so for the number of such system we 
can adopt the lower limit $N(\hbox {Cyg X-1})\ge 1$. On the other
hand, as the analysis of stellar evolution shows \cite{PSR+BH},
1 binary PSR with BH should exist per 1000 single radiopulsars
in the Galaxy. So far no such systems have been discovered, i.e.
there is an upper limit $N(\hbox{PSR+BH})/N(\hbox{BPSR})<1/700$.
Binary systems like Cyg X-1 and PSR+BH are evolutionary 
related \cite{PSR+BH} and hence put bounds on BH formation
parameters as shown in Fig. \ref{cyg_psr}: $k_{bh}>0.5$, 
$M_*>18 M_\odot$ (or equivalently, $M_{cr}>35 M_\odot$). 

\begin{figure}
\FIG{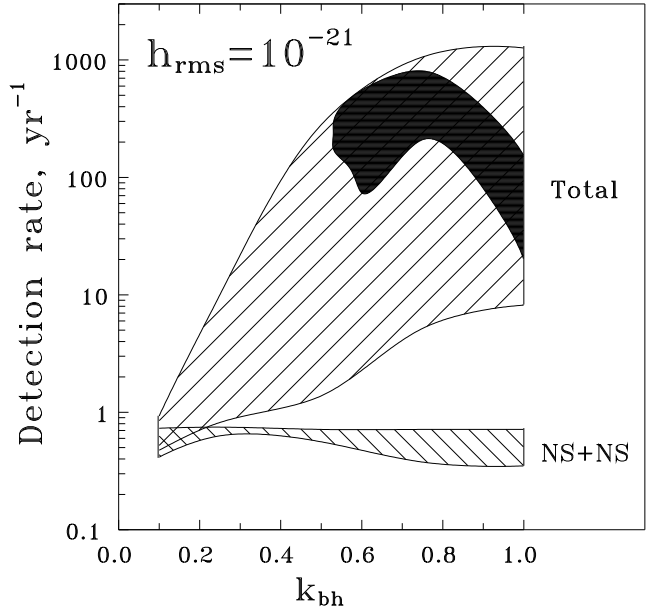}
\caption{The total merging rate of NS+NS, NS+BH, and BH+BH binaries as
would be detected by a laser interferometer with $h_{rms}=10^{-21}$ for
Lyne-Lorimer kick velocity distribution with $w_0=200-400$ km/s and BH
progenitor's masses $M_*=15-50$ M$_\odot$, for different scenarios of
binary star evolution as a function of $k_{bh}$. NS+NS mergings are
shown separately.  In all cases BH+BH mergings contribute more than
$80\%$ to the total rate. The filled ``Loch-Ness-monster-head''-like
region corrspond to BH formation parameters 
$M_*>18$ M$_\odot$ and $k_{bh}=0.5$.  }
\label{detect}
\end{figure}

Fig. \ref{detect} shows the expected total (NS+NS, NS+BH, BH+BH) and
NS+NS detection rates at $S/N=1$ level in 1-year integration on a
GW-detector with the initial laser
interferometers rms-sensitivity $h_{rms}=10^{-21}$.  All possible
combinations of BH formation parameters $M_*$, $k_{bh}$ with
Lyne-Lorimer kick velocity law with $w_0=400$ km/s were calculated for
different scenarios of stellar evolution.  It is seen from Fig.
\ref{detect} that NS+NS detection rate is $\approx 0.3-0.7$ per year
whereas BH+BH/BH+NS coalescences are parameters-dependent
and can be much more numerous.  The filled
"Loch-Ness monster"-headed 
region corresponds to the ``most realistic'' BH-formation
parameters ($M_*>18$, $k_{bh}>0.5$, $w_0=200-400$ km/s and
the ``low mass-loss'' scenario for evolution of single stars (see
\cite{PSR_deficit} for more detail).  Within this region we may expect
from $\sim$ 10 to $\sim 700$ events per year, mostly (more than
$80\%$) BH+BH coalescences.

\begin{figure}
\FIG{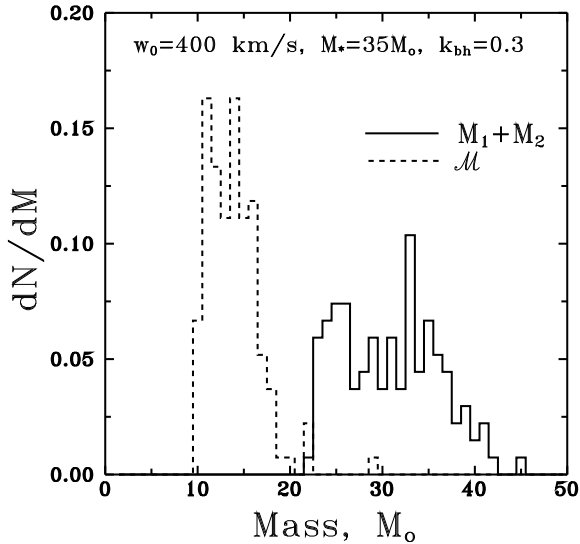}
\caption{The total mass and chirp mass distributions
for merging BH+BH binaries for $M_*=35$ M$_\odot$,
$k_{bh}=0.3$ and Lyne-Lorimer kick with $w_0=400$ km/s.
}
\label{masses}
\end{figure}

For a coalescing binary, the total mass $M=M_1+M_2$ and the chirp
mass $\cal M$ determine the final merging frequency (through the 3d
Kepler's law) and the GW-waveform amplitude, respectively.  These mass
distributions for binary BH mergings in the scenario with $M_*=35$
M$_\odot$, $k_{bh}=0.3$ and $w_0=400$ km/s are shown in Fig.
\ref{masses}. The distributions are normalized so as to give
$\int (dN/dM) dM=1$.

The results presented in this paper were obtained on the base of
astronomically well-grounded binary stellar system evolutionary
scenario.  We then may ask the question: do there exist some parameters
of the modern evolutionary scenario with which less than 1 coalescing
relativistic binary should be observed with our detector in 1-year
integration time? The answer is ``yes''  only assuming $k_{bh}<0.1$.
Then low-mass BH are formed which are subjected to the disruptive
action of both the high mass-loss during the collapse and the presumed
collapse anisotropy. These small $k_{bh}$ seem to be unrealistic
considering observed masses of BH-candidates $\sim 10 M_\odot$ 
\cite{BH_cand} (then
the progenitor's mass would be about 100 $M_\odot$ which corresponds to
implausibly high mass of 160 $M_\odot$ on the main sequence).

We conclude that for a wide range of  BH-formation
parameters ($M_{*}\approx 18-80 M_\odot$, $k_{bh}>0.5$) and
a high birth velocity of newborn collapsed stars (300-400 km/s) we
always expect $\sim 10-700$ binary merging events (mostly BH+BH pairs)
to have the signal-to-noise ratio equal unity 
during 1-year integration by the first-order GW-detectors. 
Irrespective of the absolute number of merging events (which are
subjected to at least a factor of 2 uncertainty due to unknown fraction
of baryons in binary stars), the relative number of detected BH
mergings at any detector should typically be $\sim 10$ times higher
than NS+NS coalescences.

\vskip\baselineskip
The authors acknowledge Prof. Leonid Grishchuk, Ed van den Heuvel and
Bernard Schutz for useful discussions.  KAP thanks the staff of the
Department of Physics and Astronomy of University of Wales (Cardiff) for
hospitality.  The work was partially supported by the grant of Russian
Fund for Basic Research No 95-02-06053a.

\end{document}